\documentclass[aps,showpacs,manuscript,12pt]{revtex4}
\usepackage{amssymb}
\usepackage{amsmath}
\usepackage{graphicx}

\begin{document}

\title{\textbf{On the existence of the Boltzmann-Grad\\ limit for a system of hard
smooth spheres$^{\S }$}}
\author{M. Tessarotto$^{a,b}$ and P. Nicolini$^{a,b}$} \affiliation{\
$^{a}$Department of Mathematics and Informatics, University of
Trieste, Italy, $^{b}$Consortium of Magneto-fluid-dynamics,
University of Trieste, Italy}

\begin{abstract}
Despite the progress achieved by kinetic theory, its rigorous
theoretical foundations still remain unsolved to date. This
concerns in particular the search of possible exact kinetic
equations and, specifically, the conjecture proposed by Grad
(Grad, 1972) and developed in a seminal work by Lanford (Lanford,
1974) that kinetic equations - such as the Boltzmann equation for
a gas of classical hard spheres - might result exact in an
appropriate asymptotic limit, usually denoted as Boltzmann-Grad
limit. The Lanford conjecture has actually had a profound
influence on the scientific community, giving rise to a whole line
of original research in kinetic theory and mathematical physics.
Nevertheless, several aspects of the theory remain to be addressed
and clarified. In fact, its validity has been proven for the
Boltzmann equation only at most in a weak sense, i.e., if the
Boltzmann-Grad limit is defined according to the weak *
convergence. While it is doubtful whether the result applies for
arbitrary times and for general situations (and in particular more
generally for classical systems of particles interacting via
binary forces), it remains completely unsolved the issue whether
the conjecture might be valid also in a stronger sense
(\textit{strong Lanford conjecture}). This paper will point out a
physical model providing a counter-example to the strong Lanford
conjecture, representing a straightforward generalization of the
classical model based on a gas of hard-smooth spheres. In
particular we claim that that the one-particle limit function,
defined in the sense of the strong Boltzmann-Grad limit, does not
generally satisfy the BBGKY (or Boltzmann) hierarchy. The result
is important for the theoretical foundations of kinetic theory.
\end{abstract}

\date{\today }
\maketitle



\section{Introduction: basic motivations}

Classical statistical mechanics, and in particular kinetic theory
represents, is a sense, one of the unsolved problems of classical
mechanics. In fact, although the microscopic statistical
description (MSD) of classical dynamical systems formed by
$N$-body systems is well known, a complete knowledge of their
solutions is generally not achievable. From the mathematical
viewpoint it provides an example of axiomatic approach following
from first principles and as such it must be considered as an 'ab
initio' formulation. Two equivalent treatments of MSD are known,
which are based respectively on the introduction of a phase-space
distribution function (PSDF) either on the $N$-body phase-space
$\Gamma _{N}$ or, respectively, on the 1-particle phase-space
$\Gamma _{1}.$ In the $\Gamma _{N}$ -approach the PSDF is the
so-called microscopic PSDF $f_{N}$. It follows that $f_{N}$ obeys
the Liouville equation, whose characteristics are simply the
phase-space trajectories of the same dynamical system, to be
identified with a classical $N$-body system
\cite{Grad1958,Cercignani1969}. This equation is equivalent to a
hierarchy of equations (the so-called BBGKY
hierarchy) for a suitable set of $s$-particles distribution functions ($%
f_{s}^{(N)}$), obtained letting $s=1,..,N-1$, which are uniquely
related to the corresponding PSDF. On the other hand, in the
$\Gamma _{1}-$approach the PSDF (the Klimontovich probability
density $k^{(N)}$, defined in the $\Gamma
_{1}-$space) evolves in time by means of the Klimontovich equation \cite%
{Klimontovich1958}. Also for this equation the characteristics are
just the phase-space trajectories of the $N-$body system, this
time - however - projected on the $\Gamma _{1}-$space. Therefore,
in both cases it is actually necessary to determine the
phase-space trajectories of all the particle. Hence, for classical
systems characterized by a large number of particles ($N\gg 1$),
the computational complexity (of this problem) is expected to
prevent, in general, any direct calculation of the time-evolution
either of the $N$-body or any of the the s-body distributions.
This has justified the constant efforts placed so-far for the
search of 'reduced' statistical descriptions, of which kinetic
theory (KT) is just an example. This is intended in order to
achieve efficient statistical descriptions especially suitable for
complex dynamical systems, including both gases and plasmas. \
Precisely, the primary goal of KT is the search of statistical
descriptions, either exact or in some sense approximate, whereby
the whole dynamical system is associated only to the one-particle
kinetic distribution function ($f_{1}$) defined on the
one-particle phase-space $\Gamma _{1}$, without requiring the
knowledge of the dynamics of the whole dynamical system. As a
consequence in KT-descriptions the evolution equation of the
kinetic distribution function, to be denoted as kinetic equation,
is necessarily assumed to depend functionally, in some suitable
sense, only on the same distribution function and the one-particle
dynamics. In particular, one of the most successful developments
of KT is doubtless related to the so-called 'ab initio'
approaches. These are to be intended (in contrast to heuristic or
model equations) as the KT's which are obtained deductively - by
suitable approximation schemes and assumptions - from the
corresponding exact MSD. In traditional approaches usually KT is
obtained adopting the $\Gamma _{N}$ -approach to MSD
\cite{Grad1958,Cercignani1969,Cercignani1975}. However, also the
Klimontovich method (based on the $\Gamma _{1}-$approach) can be
used \cite{Klimontovich1958}, since it is completely equivalent to
that based on the $\Gamma _{N}$ -approach \cite{Pin2001}. In all
cases KT's have the goal of determining the evolution of suitable
fluid fluid fields,\emph{\
}associated to prescribed fluids,\emph{\ }which are expressed as \emph{%
velocity moments of the kinetic distribution function } $f_{1}$
\emph{\ }and satisfy an appropriate set of fluid equations,
generally not closed, which follow from the relevant kinetic
equation. 'Ab initio' kinetic theories are - however - usually
asymptotic in character. Namely, kinetic equations are typically
satisfied only in an approximate (and asymptotic) sense and in a
finite time interval, under suitable assumptions.

\subsection{Asymptotic kinetic theories}

A well-known asymptotic kinetic equation of this type is provided
by the Boltzmann kinetic equation for a classical gas formed by
$N$ smooth rigid spheres of diameter $d$ (Grad,1958
\cite{Grad1958}), which is obtained from the exact equation of the
BBGKY hierarchy for the one-particle kinetic distribution, i.e.,
\begin{equation}
{\ {F}_{1}(\mathbf{r}_{1},\mathbf{v}_{1},t)\
f}_{1}^{(N)}{=d}^{2}{\ (N-1)C_{1}f}_{2}^{(N)},  \label{Eq.1}
\end{equation}%
where $F_{1}$ and $C_{1}$ are respectively the free-streaming operator${\ F}%
_{1}(\mathbf{r}_{1},\mathbf{v}_{1},t){\ =}\frac{\partial }{\partial t}{\ +\ }%
\mathbf{v}_{1}{\ \cdot }\frac{\partial }{\partial \mathbf{r}_{1}}$
and a suitable collision operator
\cite{Grad1958,Cercignani1969,Cercignani1975}. For definiteness,
in the remainder we adopt a dimensionless notation whereby all
relevant functions (in particular, the Newtonian particle state
${\mathbf{x}_{1}=(\mathbf{r}_{1},\mathbf{v}_{1}),}$ the time ${t}$
${,}$ the particle diameter $d$ and the volume of the
configuration space $V$) are considered non-dimensional.  The
transition from the $1$-particle equation (\ref{Eq.1}) can be
obtained by adopting a suitable asymptotic approximation and
suitable assumptions on the joint probability
densities\cite{Grad1958,Grad1972}. \ These require, in
particular, the introduction of the so-called \textit{\ rarefied gas ordering%
} (RG ordering ), to be meant both in a global and local sense,
for the
relevant physical parameters. \ More precisely, by imposing that $%
\varepsilon =1/N$ is an infinitesimal, the particle diameter $d$,
the volume $V$ of the configuration space ($\Omega $) and the
particle mass $m;$ the related global orderings are requiring to
satisfy the orderings (Grad,1958 \cite{Grad1958})

\begin{eqnarray}
d &\sim &o(\varepsilon ^{1/2}),  \nonumber \\
V &\sim &o(\varepsilon ^{0}),  \label{Global RG ordering} \\
m &\sim &o(\varepsilon ),  \nonumber \\
\eta (\mathbf{r,}t) &=&4\pi n(\mathbf{r,}t)d^{3}/3V\sim
o(\varepsilon ^{1/2}).
\end{eqnarray}%
The last ordering, in particular, prevents the number density $n(\mathbf{r,}%
t)$ from becoming so large that volume fraction $\eta
(\mathbf{r,}t)$ can be locally finite, i.e., of order
$o(\varepsilon ^{0}).$ In fact, it is
well-known that if there results locally $\eta (\mathbf{r,}t)\sim $ $%
o(\varepsilon ^{0})$ particle correlations (in particular
two-particle
correlations) may become non-negligible also on the large scale \cite%
{Grad1958,Grad1967,Tsuge1970}. In fact, these correlations, which
are not generally expected to decay rapidly in time
\cite{Grad1967}, can be also long range in character
\cite{Piasecki2007}. \ Instead, in validity of the RG ordering
defined above, uniformly in phase-space and at least in a finite
time interval $I=\left[ t_{o},t_{1}\right] ,$ with $\Delta
t=t_{1}-t_{o}$ such that $\Delta t\sim o(\varepsilon ^{0}),$ the
following conditions are assumed to be satisfied:

\begin{itemize}
\item Assumption \#1 - in $\Gamma _{s}\times I_{o1}$, the approximate (i.e.,
asymptotic) joint probability densities $f_{s}(\varepsilon )$ (for
any $s\in \mathbb{N}$ \ with $s\ll N$)\ are smooth and bounded
ordinary functions
defined in $\Gamma _{s}\times I_{o1},$ where $\Gamma _{s}$ is the $s$%
-particle phase-space;

\item Assumption \#2 - the \textit{asymptotic factorization condition }(AFC)
\begin{equation}
f_{s}(\varepsilon ,\mathbf{x}_{1}\mathbf{,..x}_{s},t)=\prod%
\limits_{i=1,s}f_{1}(\varepsilon ,\mathbf{x}_{i},t)\left[ 1+\Theta
(t-t_{o})o(\varepsilon ^{\alpha })\right]
\end{equation}%
is satisfied identically for any $s\in
\mathbb{N}
$ such that $s/N\sim o(\varepsilon )$. Here $f_{1}$ $(\varepsilon ,\mathbf{x}%
_{i},t)$ (for $i=1,s$) \ is the one-particle probability density
which satisfies the asymptotic Boltzmann equation
\begin{equation}
{\ {F}_{1}(\mathbf{r}_{1},\mathbf{v}_{1},t)\ f}_{1}(\varepsilon
,){=d}^{2}{\ NC_{1}f}_{2}(\varepsilon ,),  \label{asymptotic
Boltzmann eq}
\end{equation}
\end{itemize}

and $\Theta (t-t_{o})$ is the Heaviside theta function which vanishes for $%
t=t_{o};$

If the RG ordering and the previous assumptions hold locally
(i.e., in the
whole phase-space $\ \Gamma _{1}$ and at least in a finite time interval $%
I_{o1}\equiv \left[ t_{o},t_{1}\right] $), \ the Boltzmann equation (\ref%
{asymptotic Boltzmann eq}) is expected to be locally valid in the
same domain \cite{Shinbrot1984,Illner1986,Pulvirenti1987} at least
in an asymptotic sense. This means, introducing an arbitrary
monotonic decreasing sequence of infinitesimal parameters $\left\{
\varepsilon \right\} \equiv
\left\{ \varepsilon _{i}>0,i\in \mathbb{N}\right\} ,$ that the sequence $%
\left\{ f_{1}(\varepsilon ,\mathbf{x}_{1},t)\right\} $ defined in
terms of them is expected \textit{to} \textit{converge in a weak
(asymptotic) sense for }$\varepsilon \rightarrow 0.$ In other
words the whole domain $\Gamma
_{1}\times I_{o1}$ [existence domain of $f_{1}(\varepsilon ,\mathbf{x}%
_{1},t)]$ :

\begin{itemize}
\item the asymptotic solution $f_{1}(\varepsilon ,\mathbf{x}_{1},t)$
differs, by an error infinitesimal of order $o(\varepsilon
^{\alpha _{1}})$
with respect to the exact solution $f_{1}^{(N)}(\mathbf{x}_{1},t)$, \ being $%
\alpha _{1}$ is an appropriate strictly positive real number$.$ As
a
consequence,\textit{\ the error }$\Delta f_{1}\equiv f_{1}(\varepsilon )-$ $%
f_{1}^{(N)}$\textit{, while remaining non zero, can be taken
arbitrarily small};

\item in\ $\ \Gamma _{1}\times I_{o1}$ the Boltzmann kinetic equation
differs from the exact one-particle BBGKY equation at most by
terms of order $o(\varepsilon ^{\alpha _{2}}),$ where $\alpha
_{2}$ is an appropriate real number $0<\alpha _{2}\leq 1$ in
general different from $\alpha _{1}.$
\end{itemize}

\ Even if the rigorous proof of the global validity of the
Boltzmann equation for arbitrary initial and boundary conditions
has yet to be reached, its success in providing extremely accurate
predictions for the dynamics of rarefied gases and plasmas is well
known (see for example, Cercignani, 1969 \cite{Cercignani1969};
Frieman, 1974 \cite{Frieman1975}).

\subsection{Boltzmann-Grad limit and the Lanford conjecture}

Basic issues remain to be clarified regarding the rigorous
theoretical foundations of KT. One such problem - and the one we
want to address in this Note - refers in particular to the search
of possible exact kinetic equations and, specifically, the
conjecture suggested originally by Grad (Grad, 1972
\cite{Grad1972}) and investigated by Lanford in a seminal paper
(Lanford, 1974 \cite{Lanford1975a}; see also Frieman, 1974 \cite{Frieman1975}%
), \ that kinetic equations - such as the Boltzmann equation for a
gas of classical hard spheres - might result exact in an
appropriate asymptotic limit, denoted as \textit{Boltzmann-Grad}
(B-G) \textit{limit. }In other
words, according to this conjecture, there should exist a suitable operator%
\textit{\ }$L^{\ast }$ (denoted as B-G limit operator) such that
the limit functions $f_{s}\equiv L^{\ast }f_{s}^{(N)}$ of the
sequences $\left\{
f_{s}^{(N)}\right\} ,$ \ to be defined in terms appropriate discrete sets $%
\left\{ N_{i}\in \mathbb{N}\right\} ,$ should result exact
solutions of the equation of the Boltzmann hierarchy. The B-G
limit is customarily intended as the limiting "regime" where the
total number of particles $N$ goes to infinity, while the
configuration-space volume $V$ remains constant, the particle
diameter $d$ goes to zero in such a way that $Nd^{2}$ approaches a
finite non-zero constant and the average mass density $Nm/V=M/V$
remains finite (Grad, 1972 \cite{Grad1972}; Lanford, 1974
\cite{Lanford1975a}; Frieman, 1974 \cite{Frieman1975}), i.e.,
there results:
\begin{eqnarray}
&&\left. \frac{1}{N},d,m\rightarrow 0,\right.   \nonumber \\
&&\left. \frac{Nd^{2}}{V}\rightarrow k_{1},\right.   \label{global B-G} \\
&&\left. M=\frac{mN}{V}\rightarrow k_{2},\right.   \nonumber
\end{eqnarray}%
where $k_{i}$ ($i=1,2)$ are prescribed non-vanishing finite
constants. In the case of plasmas further analogous requirements
must be placed on the
total electric charge and current carried by each particle species \cite%
{Frieman1975,Tessarotto1999a}. In addition, the proper definition
must be made for the limit operator $L^{\ast }.$ In fact, in order
that the sequences $\left\{ f_{s}^{(N)}\right\} $ converge in some
sense it is necessary to determine their time evolution. \
According to Lanford and previous authors this can be achieved by
constructing and explicit solution of the corresponding equation
of the BBGKY hierarchy, to be represented explicitly by a
time-series expansion for each distribution $f_{s}^{(N)}$.
\cite{Lanford1975a,Alexander1975,Lanford1976,Kaniel1978,Van
Beijeren1980,Spohn1981,Lanford1981,Spohn1983,Shinbrot1984,Spohn1984,Illner1986,Pulvirenti1987,Illner1987b,Gerasimenko1990}%
\ As a consequence, it was found that $L^{\ast }$ can be defined
in the sense of weak* convergence for the sequence $\left\{
f_{s}^{(N)}\right\} $ \cite{Lanford1975a}$.$ The proof of the weak
convergence of $\left\{ f_{s}^{(N)}\right\} $ in this sense was
first reached in the seminal work of Lanford (Lanford, 1974
\cite{Lanford1975a}) who was able to prove also the
local validity of the Boltzmann equation in a finite time interval $I_{o1}=%
\left[ t_{o},t_{1}\right] $ of amplitude $\Delta t=t_{1}=t_{o}$
smaller than 1/5 and under the assumption of factorization at the
initial time for the joint-particle distribution $f_{2}$. \ Even
if this result does not suffice to justify possible meaningful
physical applications, the conjecture has actually had a profound
influence on the scientific community, giving rise to a whole line
of original research in kinetic theory and mathematical physics.
The work was later extended by other authors to include 2D and 3D
and global validity for the Boltzmann equation. \cite%
{Lanford1975a,Alexander1975,Lanford1976,Kaniel1978,Van
Beijeren1980,Spohn1981,Lanford1981,Spohn1983,Shinbrot1984,Spohn1984,Illner1986,Pulvirenti1987,Illner1987b,Gerasimenko1990}%
However, the validity of the Boltzmann equation for general
situations remains dubious. \

A key issue, however, is related to the possible validity of the
Lanford conjecture in a stronger sense, not just for the Boltzmann
equation but also for the BBGKY hierarchy itself, as following\ by
suitable definition of the
B-G limit operator $L^{\ast }$ acting on the joint probability densities $%
f_{s}^{(N)}.$ In fact, let us assume that $L^{\ast }$ is defined
in the sense of uniform convergence in phase-space of the
sequences $\left\{ f_{s}^{(N)}\right\} $ to the strong limit
functions $f_{s}\equiv L^{\ast }f_{s}^{(N)}.$ In such a case the
conjecture can be advanced that the strong limit functions $f_{s}$
belong to same functional class of $\left\{ f_{s}^{(N)}\right\} $
(\emph{strong Lanford conjecture})$.$\ \ In particular, this means
that when applying the operator $L^{\ast }$ term by
term to the equation of the BBGKY hierarchy for $f_{1}^{(N)}$ [Eq.(\ref{Eq.1}%
)]
\begin{equation}
{\ L^{\ast }F}_{{\ 1}}{\ f}_{1}^{(N)}{=L^{\ast }}\left\{ {d}^{2}{(N-1)C_{1}f}%
_{2}^{(N)}\right\} ,
\end{equation}%
the limit function $f_{1}\equiv L^{\ast }f_{1}^{(N)}$ should
satisfy the corresponding equation of the Boltzmann hierarchy

\begin{equation}
{\ F}_{{\ 1}}{\ f_{1}=k}_{1}{C_{1}f}_{2}.
\end{equation}%
For the validity of this limit equation \ it follows that $L^{\ast
}$ should commute with the streaming operator ${{F}_{{\ 1}}}$, in
the sense that it should result identically
\begin{equation}
\left[ {{L^{\ast },F}_{{\ 1}}}\right]
{f}_{1}^{(N)}(\mathbf{x}_{1},t)\equiv 0,  \label{commutation}
\end{equation}%
being $\left[ {{L^{\ast },F}_{{\ 1}}}\right] ={\ {L^{\ast }F}_{{\ 1}}{\ -}\ {%
F}_{{\ 1}}\ {L^{\ast }}}$. \ In the sequel we intend to point out,
however, that \emph{the Lanford conjecture is not generally valid
in this sense}, i.e., \ that the limit functions defined in the
strong B-G limit do not actually belong to the same functional
class of the sequences $\left\{ f_{s}^{(N)}\right\} $ and in
particular to the solutions of the Boltzmann hierarchy.
Nevertheless, weak convergence in the sense indicated above may
still be warranted. \ In order to prove the point, in this paper
we intend to propose a counter-example based on the introduction
of a modified three-dimensional hard-sphere problem.

\section{A counter-example: a modified 3D hard-sphere system}

To prove this point, we consider here a system ($S_{N}$) of $N$
partially-impenetrable hard-smooth spherical surfaces ('spheres').
The other key element of the proof is the adoption of the
Klimontovich approach. As indicated elsewhere
\cite{Pin2001,Tessarotto2008} this permits to construct an exact
explicit integral representation for the $s$-particle distribution
functions without recurring to cumbersome time-series representations \cite%
{Pin2001}. For definiteness, the system $S_{N}$ is defined by
requiring that all particles are alike with diameter $d$ and mass
$m$ and are included in a bounded and connected 3D configuration
space $\Omega $ of $\mathbb{R}^{3}$ of volume $V(\Omega ).$ The
particles of $S_{N}$ can be classified
respectively as external and internal, according to the sub-domains of the $%
S_{N}$-configuration space to which they belong, denoted
respectively as external and internal ($\Omega _{ext}$ and $\Omega
_{int}=\Omega -\Omega _{ext}$). It is assumed that the two
sub-domains are mutually inaccessible, i.e., particles cannot move
from $\Omega _{ext}$ to $\Omega _{int}$ or vice versa. \ As a
consequence the numbers of internal and external particles
(defined by the occupation numbers $N_{int}$ and $N_{ext},$with $N=N_{int}$ $%
+$ $N_{ext}$) are by assumption constant. External particles are
those whose inter-particle distances (distance between the centers
of the same spheres) is larger than (or equal) to $\ d.$ Two
particles are called mutually internal if their inter-particle
distance is smaller than (or equal) to $d.$ Internal particles
are, therefore, those such that there exists at least another
particle of $S_{N}$ with which they are mutually internal$.$ It is
required furthermore that: 1) the occupation numbers $N_{int}$ and
$N_{ext}$ are both non-zero; 2) particles and the boundary of
$\Omega $ are mutually impenetrable; 3) external particles are
impenetrable when they collide with another (external or internal)
particle; 4) two arbitrary mutually internal \ particles are, by
definition, mutually impenetrable (since the intersection between
their boundaries is always non-empty). Particles can undergo
interactions either with the boundary (unary interactions) or
among themselves (binary interactions), all assumed elastic. For
binary interactions, we distinguish between external and internal
collisions. In particular external collisions occur when two
particles - initially with an inter-particle distance large than
$d$ - touch each other. Instead, internal collisions are defined
only among mutually internal particles. \ The MSD for the system
$S_{N},$ in analogy to the customary hard-smooth sphere problem
\cite{Grad1958,Cercignani1969}, can be achieved in principle in an
elementary way by distinguishing between external and internal
subsets of phase-space, either $\Gamma _{N}$ or $\Gamma _{1}$ (see
discussion above). In particular, for example, $\Gamma _{N}^{ext}$
and $\Gamma _{N}^{int}$ are respectively the subsets of $\Gamma
_{N}$ in which particles belong respectively to the external and
internal sub-domains of the configuration space$.$ In the case of
the $\Gamma _{N}-$phase-space formulation, this implies that the
Liouville equation must be satisfied identically by the
PSDF $f_{N}$ in both subspaces ($\Gamma _{N}^{ext}$ and $\Gamma _{N}^{int}$%
). This leaves, nonetheless, a large freedom in the choice of the
initial-boundary conditions as well as the functional class of
$f_{N}$\ . In particular, due to the arbitrariness of $f_{N},$ it
is always possible to invoke the assumptions: \emph{Assumption
}$\alpha $) the PSDF $f_{N}$ results continuous in the whole set
$\Gamma _{N}$ and in particular on the boundary between external
and internal particles ($\delta \Gamma
_{N}^{ext}\equiv \delta \Gamma _{N}^{int}$); \emph{Assumption} $\beta $) $%
f_{N}$ is a smooth and bounded ordinary real function. The corresponding $%
\Gamma _{1}-$phase-space formulation for $S_{N},$ is obtained by
constructing the corresponding Klimontovich probability density.
In $\Gamma _{1}$ for external particles it reads
\begin{equation}
k^{(N_{ext})}(\mathbf{y,}t)=\frac{1}{N_{ext}}\sum_{i=1,N_{ext}}\delta (%
\mathbf{y-x}_{i}(t))\Theta _{i}(\mathbf{r,}t),
\end{equation}%
where $\mathbf{y=}\left( \mathbf{r,v}\right) $ is an arbitrary
state vector
of the one-particle phase space $\Gamma _{1}.$ Here $\mathbf{x}_{i}(t)=%
\mathbf{\chi }_{i}(\mathbf{x}_{o},t_{o},t)$ [for $i=1,N],$ denote
the phase-space trajectories of the particles of $S_{N}$ with
initial conditions $\mathbf{x}_{i}(t_{o})=\mathbf{x}_{oi}$ [for
$i=1,N$]$.$ These trajectories are assumed to be defined uniquely
in the set $\Gamma _{N}\times I,$ $I$ being a suitable bounded
time interval. and $\Theta _{i}(\mathbf{r,}t)$ is the function
$\Theta _{i}(\mathbf{r,}t)\equiv 1-\sum_{j=1,N,j\neq i}\Theta
(d-\left\vert \mathbf{r-r}_{j}(t)\right\vert ),$while $\Theta (x)$
is the Heaviside step function $\Theta (x)=\left\{
\begin{array}{l}
1\text{ if }x\geq 0 \\
0\text{ if }x<0.%
\end{array}%
\right. $ As a consequence it follows that in the subset of phase-space $%
\Gamma _{1}$ for external particles ($\Gamma _{1}^{ext}$) the
one-particle distribution function when expressed in terms of the
initial microscopic
PSPD reads $f_{1}^{(N)}(\mathbf{y,}t)=\int_{\Gamma _{N}}d\mathbf{x}%
_{o}f^{(N)}(\mathbf{x}_{o}\mathbf{,}t_{o})k^{(N_{ext})}(\mathbf{y,}t).$
Invoking the Liouville equation for $f^{(N)}$ this can be prove to
imply:
\begin{equation}
f_{1}^{(N)}(\mathbf{y,}t)=\widehat{f}_{2}^{(N)}(\mathbf{y,}t)-\widehat{I}%
_{2}^{(N)}(\mathbf{y,}t),
\end{equation}%
where $\widehat{f}_{2}^{(N)}(\mathbf{y,}t)=\int_{\Gamma _{N}}d\mathbf{x}%
f^{(N)}(\mathbf{x,}t)\delta (\mathbf{y-x}_{1}(t))$ and $\widehat{I}%
_{2}^{(N)}(\mathbf{y,}t)\equiv (N-1)\int_{\Gamma _{N}}d\mathbf{x}f^{(N)}(%
\mathbf{x,}t)\delta (\mathbf{y-x}_{1}(t))\Theta (d-\left\vert \mathbf{r-r}%
_{2}(t)\right\vert ).$ Then the following result can be reached \cite%
{Tessarotto2008}:\newline
\newline

\textbf{Theorem - Non-existence of the strong B-G limit for
$S_{N}$}\newline
\textit{\ Let us assume that there is at least a finite time interval }$%
I_{o1}=\left[ t_{o},t_{1}\right] \subseteq \mathbb{R}$\textit{\
such that the probability densities }$f_{s}^{(N)}$\textit{\
(}$s=1,2$\textit{) and their strong limit functions }
$f_{s}=L^{\ast }f_{s}^{(N)}$ \emph{(} $s=1,2$ \emph{)}
$f_{1}^{(N)}$\textit{\ is strictly positive in }$\Gamma _{1}\times
I_{o1},$\textit{\ so that there results uniformly in }$\Gamma
_{1}\times
I_{o1}$\textit{\ for }$S_{N}:$\ 1) $L^{\ast }\widehat{I}_{2}^{(N)}(\mathbf{y,%
}t))=0; $2) \textit{the strong limit function }$f_{1}(\mathbf{y},t)$\textit{%
\ reads }$f_{1}(\mathbf{y,}t)=L^{\ast }\widehat{f}_{2}^{(N)}(\mathbf{y,}t);$%
3)\textit{\ }$f_{1}(\mathbf{y},t)$\textit{\ satisfies identically
the
homogeneous equation}%
\begin{equation}
F_{1}f_{1}(\mathbf{y,}t)=0.
\end{equation}%
As consequence, we conclude that in the B-G limit the strong limit function $%
f_{1}(\mathbf{y},t)$ does not generally satisfy equation
(\ref{commutation}) and hence neither the corresponding equation
of the Boltzmann hierarchy. Hence, at least in the case of the
hard-sphere system here considered, the strong Lanford conjecture
for the BBGKY hierarchy fails.

\section{Conclusions}

In this paper the issue of the validity of the Lanford conjecture
in the sense of the strong B-G limit has been investigated. An
example case has been formulated based on the analysis of a system
of partially impenetrable
smooth-hard spheres. We have shown that if the one-particle limit function $%
f_{1}(y,t)$ is intended in the sense of the strong B-G limit it
does not generally belong to the functional class $\left\{
f_{1}\right\} $ of the solutions of the one-particle limit
equation. \textit{In other words, in such a case the limit
function is neither a solution of the corresponding equation of
the Boltzmann hierarchy nor - as a main consequence - of the
Boltzmann equation}. This result raises obviously the interesting
question whether similar conclusions can be reached for the
customary smooth-hard sphere system \cite{Grad1958,Cercignani1969}
or to more general systems of interacting particles. This problem,
together with a detailed analysis of
the approach here developed, will be discussed elsewhere \cite%
{Tessarotto2008}.


\section*{Acknowledgments}
Work developed in cooperation with the CMFD Team, Consortium for
Magneto-fluid-dynamics (Trieste University, Trieste, Italy). \
Research developed in the framework of the MIUR (Italian Ministry
of University and Research) PRIN Programme: \textit{Modelli della
teoria cinetica matematica nello studio dei sistemi complessi
nelle scienze applicate}. The support COST Action P17 (EPM,
\textit{Electromagnetic Processing of Materials}) and GNFM
(National Group of Mathematical Physics) of INDAM (Italian
National Institute for Advanced Mathematics) is acknowledged.

\section*{Notice}
$^{\S }$ contributed paper at RGD26 (Kyoto, Japan, July 2008).
\newpage



\end{document}